# Observation of quantum depletion in a nonequilibrium exciton-polariton condensate

Maciej Pieczarka[1], Eliezer Estrecho[1], Maryam Boozarjmehr[1], Olivier Bleu[2], Mark Steger[3*], Kenneth West[4], Loren N. Pfeiffer[4], David W. Snoke[3], Jesper Levinsen[2], Meera M. Parish[2], Andrew G. Truscott[5], and Elena A. Ostrovskaya[1]

[1]ARC Centre of Excellence in Future Low-Energy Electronics Technologies and Nonlinear Physics Centre, Research School of Physics, The Australian National University, Canberra, ACT 2601, Australia

[2]ARC Centre of Excellence in Future Low-Energy Electronics Technologies and School of Physics and Astronomy, Monash University, Melbourne, VIC 3800, Australia

[3]Department of Physics and Astronomy, University of Pittsburgh, Pennsylvania 15260, USA

[4]Princeton Institute for the Science and Technology of Materials (PRISM), Princeton University, Princeton, New Jersey 08544, USA

[5]Laser Physics Centre, Research School of Physics, The Australian National University, Canberra, ACT 2601, Australia

**The property of superfluidity, first discovered in liquid $^4$He, is closely related to Bose-Einstein condensation (BEC) of interacting bosons. However, even at zero temperature, when one would expect the whole bosonic quantum liquid to become condensed, a fraction of it is excited into higher momentum states via interparticle interactions and quantum fluctuations – the phenomenon of quantum depletion. Quantum depletion of weakly interacting atomic BECs in thermal equilibrium is well understood theoretically but is difficult to measure. This is even more challenging in driven-dissipative systems such as exciton-polariton condensates (photons coupled to electron-hole pairs in a semiconductor), since their nonequilibrium nature is predicted to suppress quantum depletion. Here, we observe quantum depletion of an optically trapped high-density exciton-polariton condensate by directly detecting the spectral branch of elementary excitations populated by this process. Analysis of the population of this branch in momentum space shows that quantum depletion of an exciton-polariton condensate can closely follow or strongly deviate from the equilibrium Bogoliubov theory, depending on the fraction of matter (exciton) in an exciton-polariton. Our results reveal the effects of exciton-polariton interactions beyond the mean-field description and call for a deeper understanding of the relationship between equilibrium and nonequilibrium BECs.**

The fundamental understanding of interacting cold bosonic gases was developed by N. Bogoliubov, whose theory predicted the consequences of interparticle interactions for the properties of Bose-Einstein Condensates (BECs)[1]. According to this theory, an interacting BEC[2] is characterised by a modified phonon-like dispersion of elementary excitations at long wavelengths (or short wavevectors) that explains the superfluid properties of weakly interacting BECs. The crux of the theory is the non-perturbative transformation of Bogoliubov quasiparticles, where a condensate excitation at a given momentum is expressed as a superposition of counter-propagating single-particle states[3]. The quantum fluctuations of the interacting particles in the ground state are responsible for the non-zero occupation of these elementary excitations at zero temperature, leading to the so-called quantum depletion of the condensate population. Quantum depletion was observed in weakly interacting BECs[4–6] of ultracold atoms and is the main reason for a low condensed fraction of strongly interacting superfluid $^4$He[7]. This effect is predicted to have pronounced experimental signatures, where the occupation of elementary excitation modes exhibits a distribution in momentum space $N(k)$ scaling as $k^{-4}$ for long wavevectors. This behaviour is challenging to observe in weakly interacting atomic BECs[5,6,8], because the momentum distribution is not preserved in the time-of-flight measurements and is influenced by interactions during the expansion of the condensate[9].

Exciton-polariton condensates, which are part-light part-matter bosonic condensates formed in a semiconductor microcavity, allow direct measurement of their momentum space distribution and excitations through the cavity photoluminescence signal. Each exciton polariton (or polariton) is a strongly coupled quantum well exciton and a cavity photon. Due to the finite lifetime of the confined photon state, the polariton eventually decays and a photon escapes the microcavity retaining the energy and momentum of the polariton[10]. The finite lifetime results in the inherent driven-dissipative nature of exciton-polariton condensates, where the system needs constant pumping to maintain the condensate population, hence the steady state is reached based on a balance between driving and dissipation. In addition, the finite lifetime often prevents full thermalisation of the condensate, which is typically manifested by a macroscopic occupation of several single-particle energy states rather than a single ground state. Furthermore, the constant energy flow in exciton-polariton condensates strongly influences the elementary excitation spectrum, which can possess a gapped mode or exhibits a flat dispersionless Goldstone mode at the low energy limit, gradually recovering the textbook Bogoliubov dispersion at longer wavevectors[11]. Nonetheless, exciton-polariton condensates preserve superfluid properties[12,13], although their features are currently understood as a rigid state under continuous coherent driving[14].

The striking feature of the Bogoliubov excitation spectrum is the appearance of two branches which are positive (normal branch – NB) and negative (ghost branch – GB) with respect to the condensate energy. While occupation of the NB occurs via the process of thermal excitation (thermal depletion), the GB is populated solely by the quantum depletion process, thus its appearance in the photoluminescence spectrum can serve as a direct probe of beyond mean-field effects (quantum fluctuations) in an exciton-polariton condensate[15]. Quite notably, the quantum effects of polariton-polariton interactions have been recently demonstrated, but at the single-particle level, through correlation experiments on strongly confined exciton-polaritons[16–18]. There have been several attempts to measure the full excitation spectrum of exciton-polariton condensates including populating both excitation branches using resonant pump-probe schemes[19–21] and intense incoherent driving[22,23]. However, none of these approaches created the condensates in a spontaneous, steady-state configuration. More importantly, the population of the GB was forced by scattering on defects[22] or by an additional resonant laser beam[19], hindering the direct observation of interaction-driven many-body quantum effects.

The linear Bogoliubov spectrum of a spontaneously created high-density exciton-polariton condensate has been observed, but only the positive energy NB photoluminescence was visible in the momentum space and its population was naturally dominated by thermal excitations of the condensate[24], which scaled as $k^{-2}$ in the low wavevector limit. Recently, the problem of reduced visibility of the GB was tackled theoretically[15,25], showing that nonequilibrium effects can suppress the quantum depletion and its signatures in the photoluminescence spectrum of an exciton-polariton condensate.

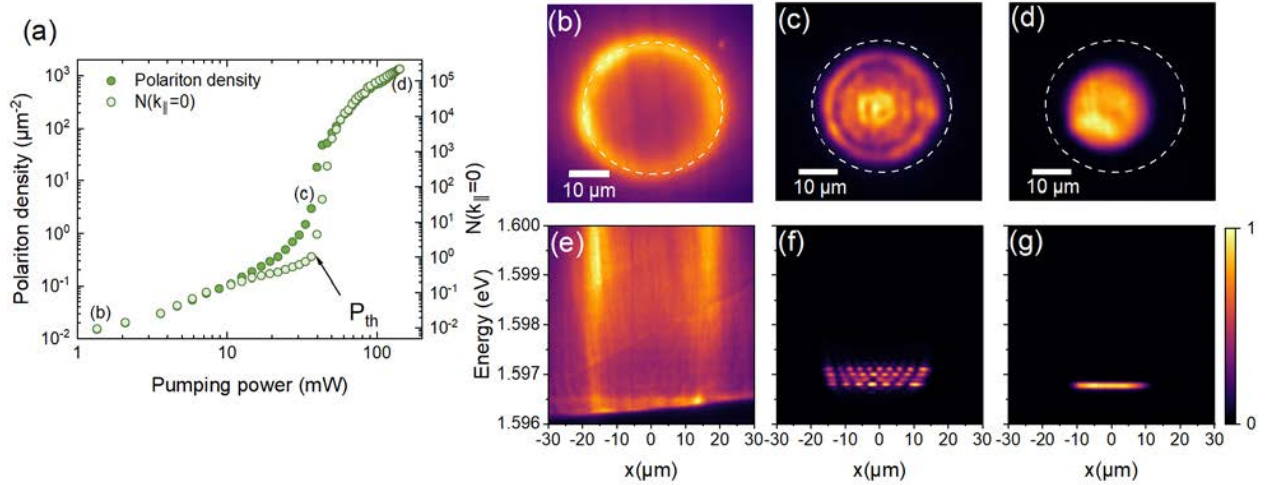

**Figure 1 High-density single-mode condensation. (a)** Pump power dependence of the total mean polariton density (filled circles) and the occupation number of the $k_\parallel = 0$ state (empty circles), both measured inside the trap. Ground state condensation threshold $P_{th}$ is indicated with an arrow. **(b-d)** Energy integrated, real-space images of exciton-polariton luminescence at the three density regimes marked in panel **(a)** shown together with **(e-g)** the corresponding real-space spectra taken in the middle of each image. **(b,e)** Images far below condensation threshold, $n < 0.02\ \mu m^{-2}$. **(c, f)** Multimode condensation in the intermediate regime, $n = 18\ \mu m^{-2}$. **(d,g)** High-density, interaction dominated single-mode regime, $n = 1340\ \mu m^{-2}$. Dashed circle in **(b-d)** represents the shape of the ring excitation. The colour scale is linear. Data is presented for the photonic detuning, $\Delta$ = -3.7 meV (see Supplementary Information for the excitonic detuning data).

In this work, we create a steady-state high-density condensate of long-lifetime exciton polaritons in an ultrahigh-quality GaAs-based microcavity, and observe the direct manifestation of quantum depletion in the condensate excitation spectra (see Methods). This is achieved by an optical excitation scheme, where a nonresonant pump laser beam with an annular spatial distribution[26,27] creates a similarly shaped distribution of incoherent excitonic reservoir particles that provides both gain and a trapping potential for the exciton polaritons[28]. At a sufficiently high pump power, a single-mode condensate in the interaction-dominated Thomas-Fermi regime[27] forms inside the trap away from the pump-induced potential barrier. The spatial separation of the pump and the condensate allows us to filter out the photoluminescence (see Methods) originating from regions with significant overlap with the reservoir and analyse the emission of the condensate and its excitations. In addition, the single-mode condensate ensures that we are dealing with a simple many-body macroscopically occupied ground state of the pump-induced trap. To test both the equilibrium and nonequilibrium features of the exciton-polariton condensate, we study the system at various values of detuning between the cavity photon and the exciton energies $\Delta = E_c - E_x$. A wide range of detuning values is accessible via the cavity wedge (see Methods), and reflects either a more matter-like (excitonic) or a light-like (photonic) nature of the polaritons at positive or negative detunings, respectively. In what follows, we present full datasets for two representative values of detuning: positive ($\Delta = +1.8$ meV) and negative ($\Delta = -3.7$ meV). The Hopfield coefficients[10] determining the exciton (matter) fraction of the polariton at these detunings are $|X|^2 \approx 0.56$ and $|X|^2 \approx 0.39$ at $k_\parallel = 0$, respectively. The excitonic or photonic nature of exciton-polaritons affects their ability to achieve thermal equilibrium[29,30].

In our experiment, we utilise a continuous wave (CW) laser and create a steady state polariton condensate. Nevertheless, condensation features are very similar to those observed in the pulsed excitation regime[27]. The typical condensate spectra are presented in Fig. 1 together with the dependency of the polariton density and the ground state occupancy on pump power (Fig. 1a). Below the condensation threshold, laser-induced high-energy excitons relax and form exciton polaritons with large wavevectors and energies, which create a local potential barrier and dominate the photoluminescence under these low-power excitation conditions, Figs. 1b, 1e. With the increase of pumping power, exciton polaritons increase in density until they condense into a fragmented condensate with the macroscopic occupation of several modes of the trap, see Figs. 1c, 1f. This fragmentation is due to the weak polariton-polariton interaction and inefficient phonon-mediated relaxation processes at intermediate densities[31,32]. Multimode condensation manifests itself in a smooth nonlinear rise in polariton density at pumping powers slightly before single mode condensation sets in, see Fig. 1a. After reaching the threshold at large polariton densities, stimulated scattering and efficient energy relaxation drive the condensation towards a single-mode ground-state condensate with a spatially homogeneous density distribution and well-defined energy, as seen in Figures 1d, 1g. In this interaction-dominated Thomas-Fermi regime, we can reliably measure the spectrum of condensate excitations, and perform a direct comparison with the Bogoliubov theory.

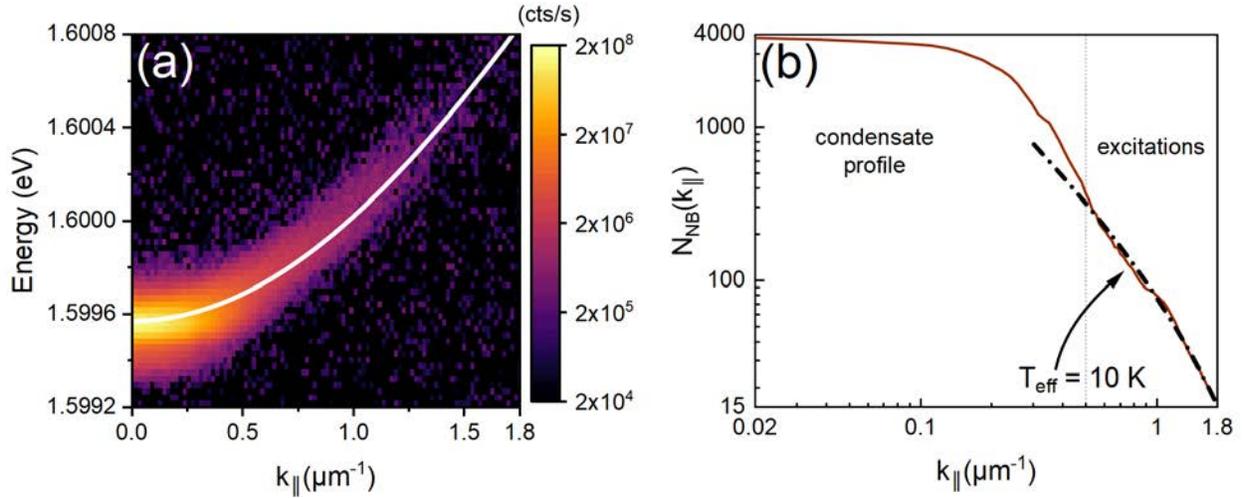

**Figure 2 Momentum space photoluminescence of a polariton condensate. (a)** Momentum-space spectrum of the exciton-polariton condensate at the excitonic detuning ($\Delta = +1.8\ meV$), where excitations can be detected together with the strong signal from the $k_\| = 0$ condensate for $n = 70\ \mu m^{-2}$. Solid line is the single-particle (non-interacting) polariton dispersion. The colour scale is logarithmic. **(b)** Extracted momentum space occupation of the condensate and the thermal excitations in the normal branch in log-log scale. Dashed-dotted line indicates the profile of thermal excitations fit with the effective temperature $T_{eff} \approx 10\ K$.

In order to investigate the excitations of the steady state exciton-polariton condensate, we measure the energy-resolved far-field emission of the condensate, which corresponds to the momentum- and energy-resolved distribution of exciton polaritons. Figure 2a presents the excitation spectrum of a single-mode condensate at intermediate densities, where the excitation spectrum is indistinguishable from the single-particle dispersion, i.e. the condensate interaction energy is too small to modify considerably the Bogoliubov spectrum within the linewidth of the photoluminescence signal. The result of the momentum space integration is presented in Fig. 2b in double logarithmic scale. One can distinguish the condensate profile (the strongest signal located near $k_\| = 0$, where $k_\|$ is the polariton momentum in the plane of the quantum well) from the thermal excitations, which are visible at longer wavevectors and are fitted with a thermal distribution, Fig. 2b (see Methods). The condensate at $k_\| = 0$ has a finite width in momentum space, which is a consequence of the spatial confinement in an optical trap[24,28].

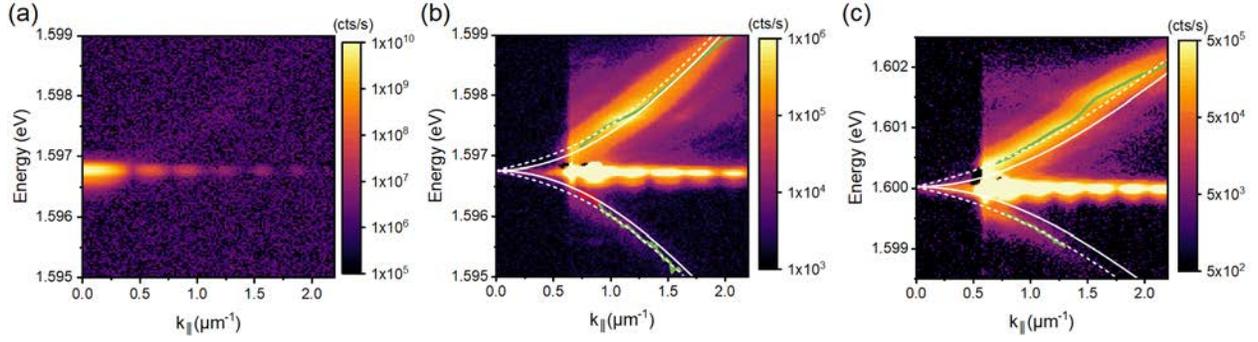

**Figure 3 Measurement of the renormalized excitation spectrum. (a)** Full photoluminescence of the high-density exciton-polariton condensate at the photonic detuning ($\Delta = -3.7\ meV$) and $n \approx 1340\ \mu m^{-2}$. **(b)** Excitation spectrum of the polariton condensate in **(a)** measured with a momentum edge filter covering the signal up to $k_\parallel \approx 0.55\ \mu m^{-1}$. **(c)** Excitation spectrum measured at the excitonic detuning ($\Delta = +1.8\ meV$) and $n \approx 1848\ \mu m^{-2}$. Color scales are logarithmic and images **(b,c)** are saturated. Green solid lines indicate extracted spectral positions of the branches. Solid white lines represent single-particle dispersions and dashed lines correspond to the renormalized Bogoliubov dispersions fitted to the data.

The momentum spectra differ significantly in the high-density regime around $n \sim 10^3\ \mu m^{-2}$, where an exceptionally strong signal from the ground state exceeds all other contributions in the spectrum within the dynamical range of the CCD camera, see Fig. 3a. The condensate emission also reveals a characteristic Airy pattern due to diffraction of the condensate photoluminescence on a circular aperture we use as a real space filter to block the emission from the trap barrier (see Methods). To reveal the much weaker signal of the excitation branches, we impose an edge filter in momentum space, blocking the strongest contribution to the signal up to about 0.55 $\mu m^{-1}$, Fig. 3b and 3c. Remarkably, the signal from the two branches becomes clear and distinguishable, despite the strong contribution of the diffracted condensate photoluminescence. The appearance of the GB signal is evidence of quantum depletion of the exciton-polariton condensate, being a consequence of quantum fluctuations in the steady state.

Fitting the excitation branches dispersions, see Fig. 3b, 3c, with Eq. 2 (see Methods and Supplementary Information) allows us to extract the chemical potential $\mu$ of the condensate as a function of the condensate density, which follows closely the expected linear dependence $\mu = gn$, where $g$ is the polariton-polariton interaction strength and $n$ is the measured polariton density, see Fig. 4a. This fitting procedure can be applied to the experimental data at various detunings, where the GB signal and the dispersion change of the excitation branches are observable. The extracted values of the polariton-polariton interaction strengths, normalised to the number of quantum wells, is presented in Fig. 4b, together with the values obtained under pulsed excitation and at larger densities ($n{\sim}10^4\ \mu m^{-2}$) in the Thomas-Fermi regime[27]. One observes an excellent agreement for all experimental values showing the theoretically predicted quadratic dependence on the exciton fraction: $g \propto |X|^4 g_X$, where $g_X$ is the exciton-exciton interaction constant[10,27], which in our case is $g_X = 13.5 \pm 0.6\ \mu eV \mu m^2$. We note that, in contrast to Ref. [27], the direct measurement of the interaction constant from the blueshift of the Thomas-Fermi condensate energy in the CW regime is not possible because of the incomplete depletion of the reservoir, which effectively lifts the zero-point energy of the optically-induced potential (see Supplementary Information). The measurement of the interaction strength presented here is therefore methodologically different and independent from that in Ref [27].

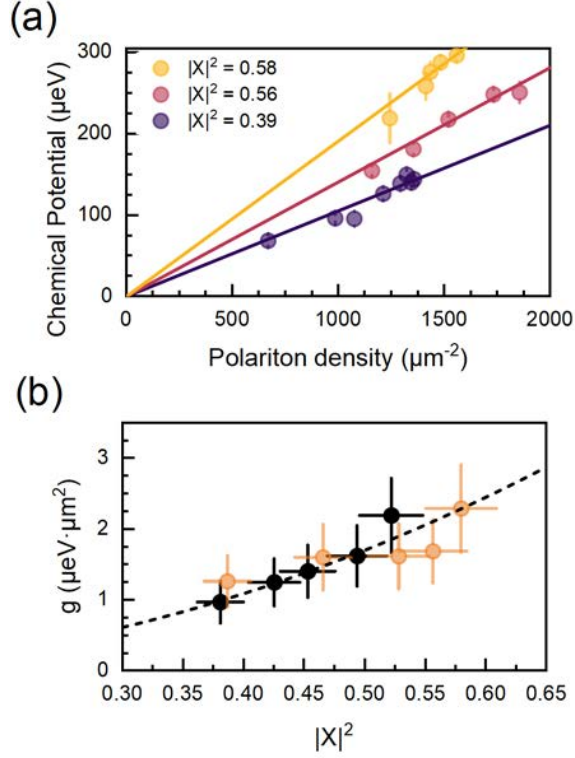

**Figure 4 Polariton-polariton interactions. (a)** Extracted condensate energies (chemical potential $\mu = gn$) from Bogoliubov dispersion fitting for three different values of detuning (other detuning values are not presented for clarity of the plot). **(b)** Summary of extracted interaction strengths per single quantum well as a function of the excitonic Hopfield coefficient (exciton fraction). Orange dots represent data collected from fitting the GB dispersion, as described in the text. The data is in excellent agreement with values obtained from an independent experiment in a pulsed excitation regime[27] (black dots). Fit to the data using a quadratic function of $|X|^2$ is shown with a dashed line.

The essential information on the mechanisms populating the elementary excitation branches, summarised in Fig. 5, is contained in the characteristic momentum occupation distributions $N(k_\parallel)$. The employed filtering technique allows us to measure these distributions in the long wavevector range, i.e. $k_\parallel \xi > 1$, where $\xi = \frac{\hbar}{\sqrt{2mgn}}$ is the healing length and $m$ is the polariton effective mass (see Supplementary Information). In this wavevector range, the NB occupation is characterised by nonequilibrium features at both probed values of the exciton-photon detuning, see Figs. 5b and 5c. At densities $n < 10^3 \ \mu m^{-2}$, $N_{NB}$ displays thermal-like distributions at higher momenta for the excitonic detuning. However, at the largest probed densities, thermal excitations are reduced, leading to a population of excited trap states for both photonic and excitonic polaritons, Fig. 5b and 5c. This is caused by the interplay of the different mechanisms populating the NB, as schematically depicted in Fig. 5a. In addition to thermal and quantum depletion of the condensate, the inefficient energy relaxation of polaritons naturally leads to an occupation of high-$k_\parallel$ states (see Supplementary Information). These high-energy polaritons originate from the potential barrier region (the reservoir) and are more prevalent for more photonic polaritons where the interactions and thermalisation are much weaker[10,29,30]. The peak around $k_\parallel = 2 \ \mu m^{-1}$ observed at the excitonic detuning in Fig. 5b originates from high-energy states on top of the pump region, which spread all over the area of the trap, and are similar to a high-energy state observed in a single spot excitation of a high-quality sample[33]. Additionally, polariton buildup and accumulation at high-energy states may be linked to the polariton-to-reservoir upconversion mechanism, which has been observed at strong coherent pumping fields in multiple independent experiments and is a source of the incoherent reservoir generation within the trap[17,18,34–36].

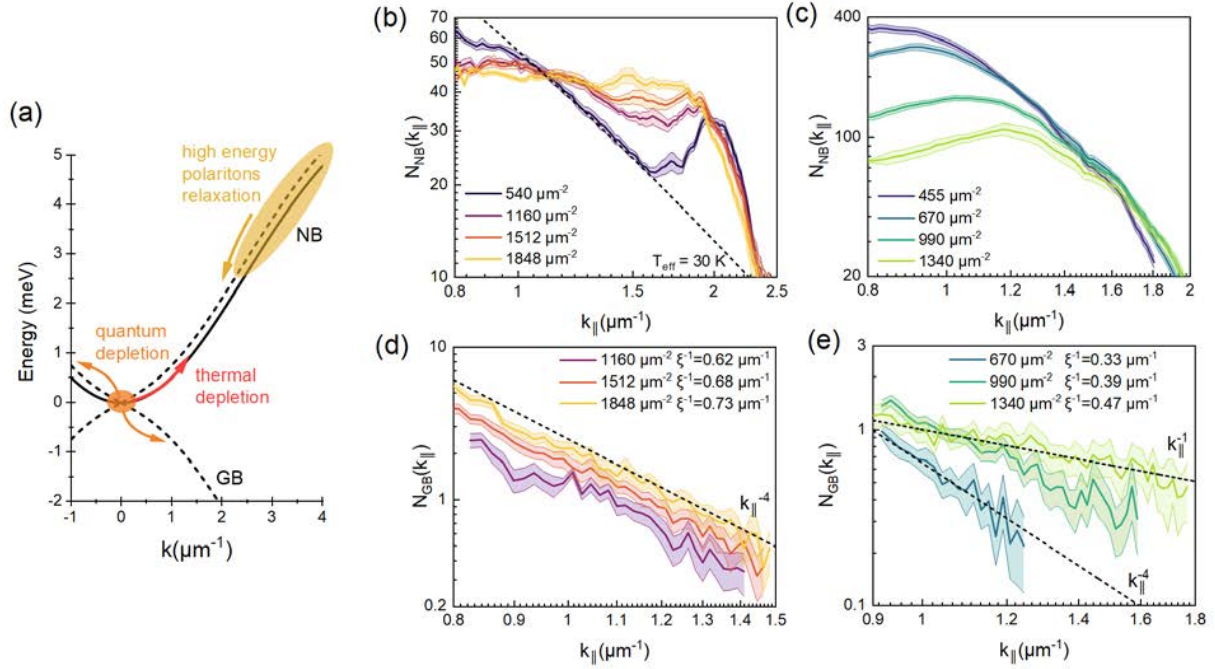

**Figure 5 Depletion mechanisms - analysis of the extracted momentum space distributions.** **(a)** Schematics showing the renormalised dispersions of the condensate excitations and possible occupation mechanisms due to quantum depletion, thermal depletion and relaxation of high-energy polaritons. **(b,c)** Normal branch occupation for high-density polariton condensates at the **(b)** excitonic ($\Delta = +1.8\ meV$) and **(c)** photonic ($\Delta = -3.7\ meV$) detunings. Dashed line in **(b)** is a guide to the eye reference of a thermal distribution of excitations at an effective temperature $T_{eff} \approx 30\ K$. High-$k_{\|}$ polaritons are more pronounced at higher densities for the photonic detuning. **(d,e)** Ghost branch occupancies for **(d)** excitonic and **(e)** photonic detunings. For more excitonic polaritons, the GB occupation behaves according to the equilibrium theory, following the asymptotic power-law decay at larger momenta. At the photonic detuning, the GB occupation shows a deviation from the Bogoliubov theory. Similar behavior is observed at other values of excitonic or photonic detuning, as shown in Supplementary Information. Plots **(b-e)** are in log-log scale and the calculated crossover wavevectors $k_\xi = \xi^{-1}$ are indicated in the legends of panels **(d,e)**.

In contrast to the NB, the GB is populated solely by quantum fluctuations of the high-density polariton condensate. Therefore, one expects to observe clear signatures in the momentum space distribution as predicted by the Bogoliubov theory. Indeed, the occupation distribution shows a $k^{-4}$ decay for the excitonic detuning, see Fig. 5d. This is in agreement with the Bogoliubov theory prediction of the asymptotic behaviour at large wavevectors and suggests the equilibrium-like character of the condensate quantum depletion despite the deviations caused by nonequilibrium effects in the NB distributions. Additionally, we observe a predicted quadratic increase of the GB occupation as a function of the condensate density (see Methods and Supplementary Information). The most remarkable behaviour of the GB occupation distributions is observed at the photonic detuning, Fig. 5e. Here, in spite of the rising population density (and pump power), we observe a gradual transition from the $k^{-4}$ distribution to a plateau approaching $k^{-1}$. The deviations from the equilibrium distributions have been predicted by nonequilibrium theories[37–39], indicating that at photonic detunings polariton condensate fluctuations are mediated mostly by reservoir-condensate interactions, and polariton-polariton interactions do not play a major role. This assertion is supported by the fact that the polariton-polariton interaction is weaker for more photonic polaritons (Fig. 4b), and that the reservoir density extracted from our measurements grows with pump power, with the condensate fraction, defined as $\rho = n/(n + n_R)$, reaching maximum values of $\rho \approx 0.5$ (see Supplementary Information). Furthermore, as discussed in the Supplementary Information, condensates at photonic detunings might be subject to reservoir-driven large fluctuations, which could lead to departure from the Bogoliubov prediction for the GB occupation[39].

It is important to note that the unusual momentum distributions $N_{GB}(k)$ observed at photonic detunings do not originate from the nonparabolic dispersion of polaritons. Taking into account the full polariton dispersion rather than the parabolic effective-mass approximation, leads to a deviation from the $k^{-4}$ behaviour of the population numbers, which is small within the experimentally accessed range of momenta (see Supplementary Information and Methods).

To summarise, in our system, we observe a crossover from the near-equilibrium regime of quantum fluctuations at the excitonic detuning ($\Delta = +1.8$ meV) to a fully nonequilibrium regime at the photonic detuning ($\Delta = -3.7$ meV), where reservoir fluctuations might play a critical role. Our experimental findings, being a first direct probe of quantum fluctuations in a many-body driven-dissipative BEC, are beyond the current theoretical understanding of nonequilibrium condensates. Thus, they call for further development of a theory describing the quantum depletion of BECs in the crossover from thermal equilibrium to far-beyond equilibrium conditions. Furthermore, in the regime where the equilibrium Bogoliubov theory of quantum depletion does apply, our experiment paves the way for the measurement of the Tan's contact[5,40] for exciton-polariton condensates (see Methods). This, in turn, will allow us to test whether exciton-polariton condensates exhibit universal thermodynamical properties of a system with contact interactions[41].

## Acknowledgements

MP would like to acknowledge useful discussions with Ryo Hanai. The work was supported by the Australian Research Council (ARC) through the Centre of Excellence Grant No. CE170100039.

## Methods

### Experiment

The sample used in the experiment is a high-quality factor GaAs-based microcavity characterised by a long cavity photon lifetime exceeding 100 ps[42]. The $3\lambda/2$ cavity consists of distributed Bragg reflectors with 32 (top) and 40 (bottom) pairs of alternating $Al_{0.2}Ga_{0.8}As/AlAs$ layers and an active region of 12 GaAs/AlAs quantum wells of 7 nm nominal thickness positioned in three groups at the maxima of the confined photon field. The normal mode anticrossing (Rabi splitting) is measured to be about $\hbar\Omega = 15.9 \pm 0.1$ meV[43], the exciton resonance energy is $E_X = 1.6062$ eV, and the cavity photon effective mass is about $3.6 \times 10^{-5}\ m_0$, where $m_0$ is the free electron mass[27,44]. In all experiments the microcavity is kept in a continuous flow helium cryostat, ensuring the sample temperature of 7-8 K.

The nonresonant excitation in the experiment was provided by a single-mode continuous-wave Ti:Sapphire laser tuned to a reflectivity minimum of the microcavity (around 719 nm) for efficient photon absorption in the quantum wells. To minimise the thermal heating of the sample, the laser pump was chopped with an acoustic optical modulator at 10 kHz and 5% duty cycle. The ring-shaped excitation profile is created by utilising an axicon lens in a confocal configuration between two imaging lenses that produces a hollow beam[45] when re-imaged onto the sample surface via a microscope objective of NA = 0.5. The same objective collects the photoluminescence from the sample. The imaging setup is composed of four lenses in confocal configuration for measurements of near and far-field emission image planes[26]. The image filtering is performed in the intermediate conjugate planes with an optical iris in real-space and a movable razor-blade edge in momentum space. The filtered signal is then imaged onto the monochromator slit and dispersed by a grating, allowing measurement of spatial and momentum spectra of exciton-polariton photoluminescence. The signal is recorded by a high-efficiency EMCCD camera.

**Extraction of the occupation numbers in momentum space**

Calculation of the momentum distribution is obtained based on the integration of the signal and taking into account the local density of states in momentum space[26]. The collection efficiency of the experimental setup was calibrated with a reference laser tuned to the emission wavelength of polaritons and is expressed as $\eta$. The mean number of polaritons recorded within a single pixel row on a CCD camera representing a wavevector $k_i$ is calculated from the photon count rate $\frac{dN_{ph}(k_i)}{dt}$:

$$N_{pol}(k_i) = \eta \frac{dN_{ph}(k_i)}{dt} \tau_{LP}(k_i),$$

where $\tau_{LP}$ is the polariton lifetime calculated based on the Hopfield coefficients and cavity photon lifetime[10,26,46]. The occupation number of polaritons at a given $k_i$ state is calculated taking into account the number of states subtended by a pixel at $k_i$ position in cylindrical coordinates $N_{st}(k_i) = k_i \Delta k_i \Delta \varphi_i \cdot \left(\frac{4\pi^2}{A}\right)^{-1}$:

$$N(k_i) = \frac{N_{pol}(k_i)}{N_{st}(k_i)} = \frac{4\pi^2 \eta}{2 k_i \Delta \varphi_i \Delta k_i DA} \frac{dN_{ph}(k_i)}{dt} \tau_{LP}(k_i).$$

The formula takes into account the volume of a single state in momentum space, spin degeneracy of 2, and the momentum space volume subtended by a single pixel. The duty cycle of the acousto-optic modulator (AOM) is denoted by $D$, and the real space filter area is denoted by $A$. A detailed derivation of the integration formulas is given elsewhere[26].

**Analysis of the Bogoliubov excitation branches**

One can diagonalise a simplified Hamiltonian of a Bose-Einstein condensate of weakly interacting bosons using quasiparticle operators at a given wavevector[47] $k$: $\hat{b}_k = u_k \hat{a}_k + v_{-k} \hat{a}_{-k}^\dagger$, which is a linear combination of creation and annihilation single-particle operators for a boson at given wavevector, $\hat{a}_k^\dagger$, $\hat{a}_{-k}$ and the amplitudes are expressed as:

$$u_k, v_{-k} = \pm \frac{1}{\sqrt{2\epsilon(k)}} \sqrt{E(k) + gn \pm \epsilon(k)}, \quad (1)$$

where $\epsilon(k)$ is the Bogoliubov dispersion in equilibrium

$$\epsilon(k) = \sqrt{E(k)(E(k) + 2gn)}, \quad (2)$$

and, $E(k) = \hbar^2 k^2/(2m)$. Occupation of the single-particle excited states is expressed as $\langle \hat{a}_k^\dagger \hat{a}_k \rangle = (|u_k|^2 + |v_{-k}|^2) \langle \hat{b}_k^\dagger \hat{b}_k \rangle + |v_{-k}|^2$. Here, the first term, proportional to $\langle \hat{b}_k^\dagger \hat{b}_k \rangle = 1/[\exp(\epsilon(k)/k_b T) - 1]$, is describing the thermal depletion of the condensate. At zero temperature, $T = 0$, there is no thermal occupation of Bogoliubov quasiparticles $\langle \hat{b}_k^\dagger \hat{b}_k \rangle = 0$, and the nonzero occupation of single-particle states is due to quantum depletion $\langle \hat{a}_k^\dagger \hat{a}_k \rangle = |v_{-k}|^2$. The GB states are populated due to quantum depletion, therefore the GB occupation can be expressed as $N_{GB}(k) = |v_{-k}|^2$.

Taking the asymptotic behaviour of $N_{GB}$, one finds that in the low wavenumber regime $N_{GB}(k) \xrightarrow{k\xi \ll 1} \frac{\sqrt{mgn}}{2\hbar k} \propto k^{-1}$, which was inaccessible in our experiment, and in the opposite regime $N_{GB}(k) \xrightarrow{k\xi \gg 1} \frac{m^2 g^2 n^2}{\hbar^4 k^4} \propto k^{-4}$ (see details in Supplementary Information). The asymptotic value $\lim_{k\to\infty} N(k) k^4 = C$ is referred to as Tan's contact[40], a universal quantity relating contact interactions to the thermodynamics of a many-body system. The contact in the equilibrium theory depends quadratically on peak density $C \propto n^2$, in agreement with the values extracted from the experiment (see Supplementary Information).

In the case of driven-dissipative systems, the Bogoliubov dispersion is modified and, after adiabatic elimination of the reservoir modes, the dispersion can be expressed in the analytical form[11]:

$$\epsilon_{neq}(k) = -i\frac{\hbar\Gamma}{2} + \sqrt{\epsilon^2(k) - \left(\frac{\hbar\Gamma}{2}\right)^2}, \qquad (3)$$

where the nonequilibrium relaxation parameter $\Gamma \xrightarrow{n\to\infty} \gamma_{LP}$ and $\gamma_{LP}$ represents the polariton decay rate. In our case, the polariton lifetime exceeds 100 ps[42], so $\hbar\Gamma \approx 3-5\ \mu eV$, which has a negligible contribution to the fitted dispersion at large wavevectors measured in this work. Therefore, for the purpose of the analysis, we use an approximation of equilibrium dispersion $\epsilon(k)$, Eq. (2). The fitting of the dispersions of elementary excitations is performed using a single fitting parameter, the condensate interaction energy $\mu = gn$, and by taking into account the experimentally measured polariton dispersion $E(k) = E_{LP}(k) - E_{LP}(0)$, where, $E_{LP}(k) = \frac{1}{2}\left(E_X + E_C(k) - \sqrt{(\hbar\Omega)^2 + \Delta(k)^2}\right)$.

# Supplementary Information:

# Observation of quantum depletion in a nonequilibrium exciton-polariton condensate

Maciej Pieczarka[1], Eliezer Estrecho[1], Maryam Boozarjmehr[1], Olivier Bleu[2],
Mark Steger[3*], Kenneth West[4], Loren N. Pfeiffer[4], David W. Snoke[3],
Jesper Levinsen[2], Meera M. Parish[2], Andrew G. Truscott[5],
and Elena A. Ostrovskaya[1]

[1]ARC Centre of Excellence in Future Low-Energy Electronics Technologies and Nonlinear Physics Centre, Research School of Physics, The Australian National University, Canberra, ACT 2601, Australia

[2]ARC Centre of Excellence in Future Low-Energy Electronics Technologies and School of Physics and Astronomy, Monash University, Melbourne, VIC 3800, Australia

[3]Department of Physics and Astronomy, University of Pittsburgh, Pennsylvania 15260, USA

[4]Princeton Institute for the Science and Technology of Materials (PRISM), Princeton University, Princeton, New Jersey 08544, USA

[5]Laser Physics Centre, Research School of Physics, The Australian National University, Canberra, ACT 2601, Australia

---

* Current address: National Renewable Energy Lab, Golden, Colorado 80401, USA

## 1. Condensate filtering in real space

To ensure that the collected photoluminescence (PL) originates from the condensate in the trap only and not from polaritons near the pump region, we use a real space filter as shown in Supplementary Figure 1a. The filter is an iris placed in the intermediate real space (near-field) image plane of the optical setup, centred with respect to the condensate position. The diameter is optimized to minimise the diffraction on its edges and to filter out the signal from the barrier region. All dispersion measurements presented in the main text were performed using this filtering technique. Supplementary Figure 1b shows the real-space spectrum when we combine the real space and the momentum space edge filters. Similarly to the momentum-resolved spectra of Figure 3 of the main text, the dominant emission originates from the condensate at $E \sim 1.600$ meV. More importantly, one can observe that the rest of the emission that constitute the NB and GB states are spread all over the trap and do not come from the barrier region. The signal outside the real space filter is due to diffraction.

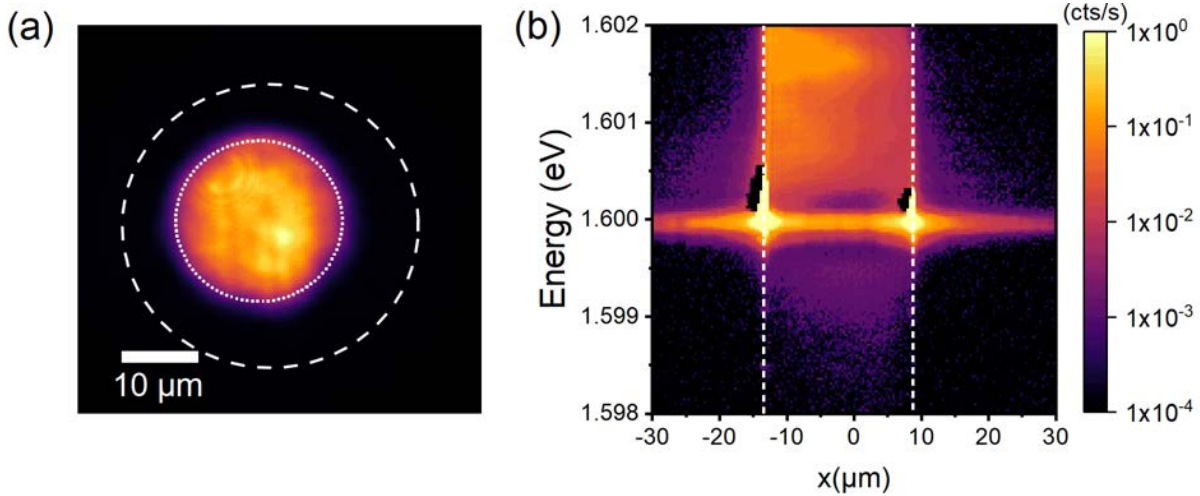

**Supplementary Figure 1 (a)** Energy integrated PL of the high-density condensate showing the edge of the real space filter (dotted circle) and the outline of the excitation profile (dashed circle). **(b)** Real space spectrum of the high-density condensate filtered in both real space (dotted lines) and momentum space (edge filter, not shown). The signal outside the real space filter is due to diffraction on the filter edges. Image is saturated and the colour scale is logarithmic.

## 2. Condensation at the excitonic detuning

Supplementary Figure 2 presents the data taken for the excitonic detuning ($\Delta = +1.8$ meV), which shows similar features to the photonic detuning case, see Figure 1 of the main text. In the power-dependent density measurements, Supplementary Figure 2a, one can observe a nonlinear growth of the density before the condensation at $k_\parallel = 0$ starts at around 70 mW. Similarly to the photonic detuning case, this is caused by the macroscopic occupation of excited states at lower pump power due to inefficient relaxation to the ground state. Additionally, there are macroscopically occupied high-energy modes located on top of the barrier[1], as can be seen in Supplementary Figures 2b, 2d. These modes coexist with the single-energy condensate inside the trap and manifest themselves in the peaked occupation at high k-vectors, see Figure 5 of the main text.

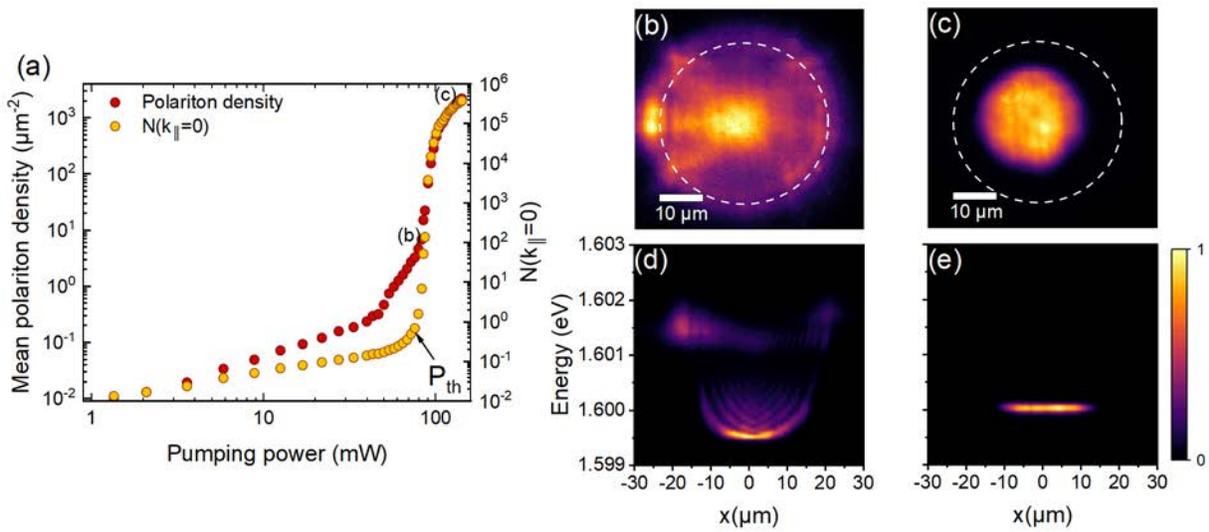

**Supplementary Figure 2 (a)** Pump power dependence of the total mean polariton density inside the trap (dark circles) and the occupation number of the $k_\parallel = 0$ state (light circles). Ground state condensation threshold $P_{th}$ is indicated with an arrow. **(b, c)** Energy integrated, real-space images of exciton-polariton photoluminescence at two density regimes marked in panel **(a)** shown together with **(d, e)** the corresponding real-space spectra taken in the middle of each image. **(b, d)** Intermediate density regime, $n \approx 7\ \mu m^{-2}$, where the high density mode is visible together with the ground state. **(c, e)** High-density, single mode condensation, $n \approx 1848\ \mu m^{-2}$.

## 3. Density-dependent blueshift of the ground state and extraction of the reservoir density

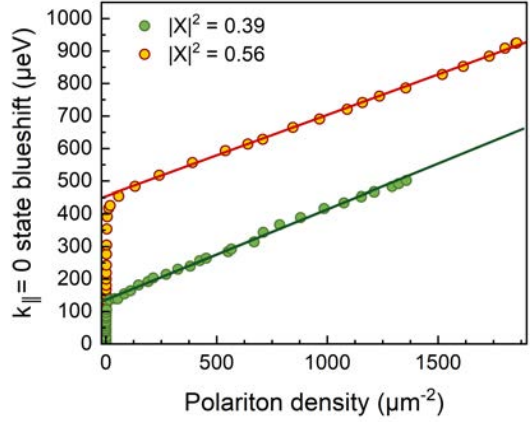

**Supplementary Figure 3** Density-dependent blueshift of the ground state extracted from the $k_\parallel = 0$ spectrum. Solid lines are linear fits to the high-density data.

Supplementary Figure 3 presents the exciton-polariton energy blueshifts extracted from the density-dependent measurements. Below the condensation threshold, at low densities, there is an anomalously large blueshift due to interaction of polaritons with the optically-injected reservoir[2,3]. At large densities above the condensation threshold, the blueshift is a linear function of the polariton density in the trap, indicating that the condensate is in the interaction-dominated Thomas-Fermi regime. We note, however, that under CW excitation the spatial depletion of the reservoir[4,8] is incomplete, which means that there is a non-negligible contribution of polariton-reservoir interactions to the blueshift within the trap. This is manifested in larger-than-expected slopes and non-zero low-density limit, $\Delta E$, of the linear dependencies of the blueshift on density, $E = sn + \Delta E$. In the cases presented here, the slopes are: $s = 0.249 \pm 0.003$ µeVµm² for the excitonic detuning and $s = 0.279 \pm 0.004$ µeVµm² for the photonic detuning. These values are about two times larger than those expected to arise due to polariton-polariton interactions and the slope is larger for the photonic detuning (smaller Hopfield coefficient), which contradicts the expected behaviour of the blueshift, based on the expression of the polariton-polariton interaction energy [8-10]:

$$\mu = gn = \frac{|X|^4}{2N_{QW}} g_X n,$$

where $|X|^2$ is the excitonic Hopfield coefficient determining the fraction of an exciton in the exciton-polariton quasiparticle, and $g_X = 6E_0 a_B^2$ is the exciton-exciton interaction constant[9], with $E_0$ and $a_B$ denoting the exciton binding energy and Bohr radius, respectively. The factor of 2 comes from the fact that the condensate is linearly polarized (i.e., has an equal mixture of two spin components), and the interaction of excitons with the opposite spin is negligible. The total interaction constant for total polariton density distributed amongst all quantum wells should be divided by $N_{QW}$.

The correct polariton-polariton interaction constants, $g$, were extracted from fitting the ghost branch of the Bogoliubov dispersion at various detunings and densities, using the interaction energy, $\mu = gn$, as a single fitting parameter (see Methods in the main text and Section 5 below). This allows us to separate the contribution of the polariton-reservoir interaction to the total blueshift shown in Supplementary Figure 3. Assuming that the total blueshift of the measured $k_\parallel = 0$ state originates from the polariton-polariton interactions and polariton-reservoir interactions: $E = gn + g_R n_R$, where $g_R = g/|X|^2$, one can extract the reservoir density, $n_R$, at a given pumping power. Here we neglect the quantum confinement effect, which would add a correction at photonic detunings and very low polariton densities below the condensation threshold. Above the condensation threshold, where the single-mode regime is

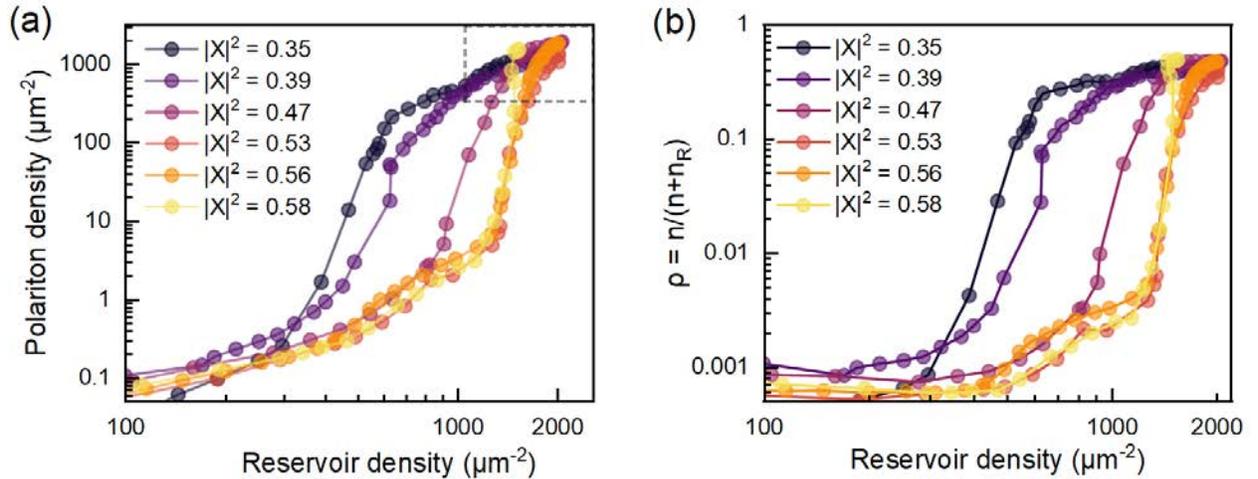

**Supplementary Figure 4 (a)** Polariton density as a function of reservoir density at different detunings (exciton fractions), extracted from a blueshift of the condensate emission energy and taking the measured polariton-polariton interaction strengths at a given detuning. The dashed rectangle indicates the density range where the GB signal is strong enough to be detected. **(b)** Condensate fraction $\rho$, calculated from the data presented in **(a)**.

achieved, this effect has negligible impact on the blueshift[3]. The measured polariton densities as a function of extracted reservoir densities are presented in Supplementary Figure 4a. One can observe increasing reservoir densities at the polariton condensation thresholds with respect to the excitonic fraction of polaritons, reaching about 1500 $\mu m^{-2}$ threshold value at excitonic detunings $|X|^2 > 0.5$. Thus, the reservoir density is non- negligible in our experiment, and the condensate fraction, defined as $\rho = n/(n + n_R)$, reaches the maximum values of around $\rho \approx 0.5$ for all detunings, see Supplementary Figure 4b. Interestingly, similar values were recently deduced by tracing the polariton condensate excitations under resonant excitation[5]. It is also important to note that the reservoir densities inside the circular trap are much smaller in comparison to excitation with a large Gaussian spot[6]. Additionally, one can point out a significant difference between the photonic and excitonic polaritons in the polariton density curves presented in Supplementary Figures 4a and 4b. At polariton densities of around $10^3 \ \mu m^{-2}$, where the GB signal was detectable, one reaches the gain saturation of stimulated scattering from reservoir at photonic detunings, where the growth of polariton density as a function of reservoir density is suppressed. This regime is not reached at excitonic detunings, even at the largest experimentally achievable polariton densities inside the trap.

The two types of behaviour displayed at photonic and excitonic dedunings can be qualitatively understood by employing a simplified rate-equation model for the polariton and reservoir density above the condensation threshold arising from the open-disspative Gross-Pitaevskii equation[10]:

$$\frac{dn}{dt} = -\gamma n + R n n_R; \quad \frac{dn_R}{dt} = -\gamma_R n_R - R n n_R + P,$$

where $n, n_R$ are the condensate and reservoir densities, $\gamma, \gamma_R$ are the respective radiative decay rates, $R$ is the rate of stimulated scattering into the condensate mode, and $P$ is the rate of pump-driven reservoir injection. For very photonic detunings ($|X|^2 \ll 0.5$), no condensate forms at the pump region ($n = 0$), so that the reservoir density at the trap barrier continues to grow lineraly with the pump rate. A non-negligible fraction of this density, $F$, is found in the centre of the trap, with the steady state defined as $n_R^0 \approx FP/\gamma_R$. The steady-state condensate density determined from the rate equations as $n^0 \approx (1 - F)\gamma_R/(FR)$ is independent on the pump rate (and reservoir density) in this regime. This is the gain saturation regime that is approached at negative (photonic) detunings in Supplementary Figure 4. In the opposite regime of excitonic detunings ($|X|^2 \gg 0.5$), the condensate first forms at the pump region, which clamps the reservoir density inside the trap at a constant value $n_R^0 \approx F\gamma/R$. However, the steady state condensate density continues to grow with the pump rate as $n^0 \approx P/F\gamma - \gamma_R/R$. As seen in Supplementary Figure 4, this behaviour manifets itself in a steepening dependence $n^0(n_R^0)$ at excitonic detunings.

## 4. Condensate spatial shape at large densities

The high-density, interaction-dominated regime is characterised by a smooth and nearly-homogeneous distribution of the ground state wavefunctions within the area defined by the optical trap. To verify this property, we extracted the real-space density distributions of condensed polaritons, which is reflected in the intensity distribution of the cavity photoluminescence taken along the real-space spectrum at the energy of the condensate, see Supplementary Figure 2e and Figure 1g in the main text. The measured profiles in the high-density regime are summarized in Supplementary Figure 5. At large densities, $n > 10^3 \ \mu m^{-2}$, the shape of the ground state wavefunction remains almost constant with further increase in density. The small modulations on the top of the condensate wavefunction is a result of local sample imperfections, reflecting the actual potential landscape of the sample.

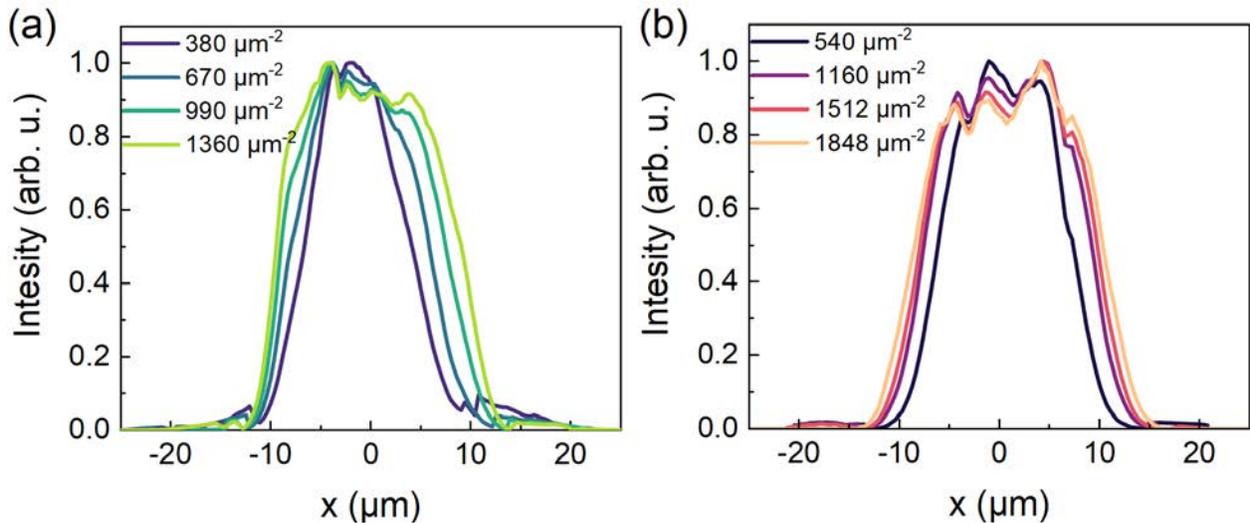

**Supplementary Figure 5** Extracted real space profiles of the condensate ground state in the high-density regime for **(a)** photonic detuning ($|X|^2 = 0.39$) and **(b)** excitonic detuning ($|X|^2 = 0.56$). Data is normalized to the maximum value.

## 5. Extraction and fitting of the excitation branches

Below we provide details of the fitting procedure of the excitation spectra presented in Figure 3 of the main text. The excitation spectra in momentum space are recorded on the CCD camera, where each pixel column corresponds to a wavevector $k_\parallel$. Examples of spectral profile at different finite wavevectors are presented in Supplementary Figures 6a,b and 6d,e. The signal is dominated by the diffracted light from the condensate, which arises from the real space filtering. On the high-energy side, one can observe the photoluminescence of the normal branch (NB), whose spectral lineshape is irregular and is composed of the occupation of many confined excited states of the optically-induced trap. The lineshape of the NB is fit with a Voigt function to extract the intensity and energy from the peak area and centre, respectively, see Supplementary Figures 6a, 6d. The low-energy peak is the signal of the ghost branch (GB), which is much weaker, as shown in the zoomed-in plots in Supplementary Figures 6b and 6e. In contrast to the NB, the lineshapes of the GB states are smooth and are fitted to Lorentzian functions for extraction of the position and intensity of the energy peaks. The extracted energy dispersion of the GB is then fitted with the Bogoliubov spectrum $\epsilon(k)$ (see main text, Methods), with the condensate interaction energy $\mu = gn$ is the only fitting parameter. Examples of the renormalized spectra fits are presented in Supplementary Figures 6c and 6f. Note that we only fit the GB dispersion due to the clarity of this data, as the GB states are populated only via quantum depletion. This is unlike the NB that has additional contributions from high-energy states coming from the top of the barrier, which results in an extracted dispersion that deviates from the Bogoliubov prediction, see Figure 3 in the main text.

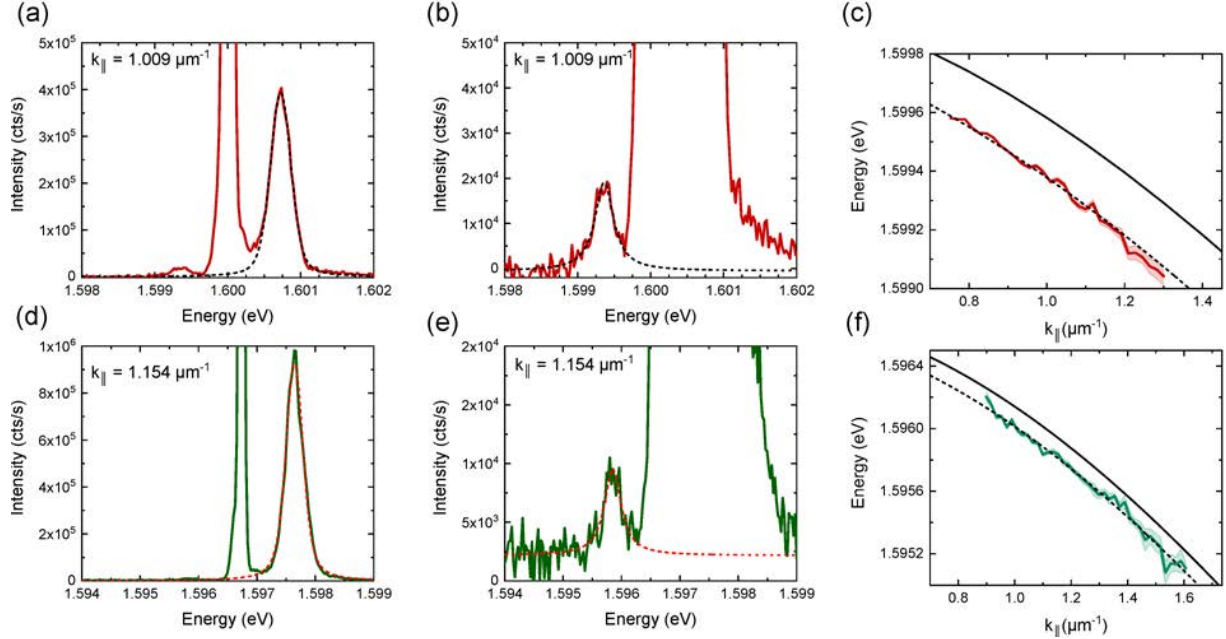

**Supplementary Figure 6** Examples of fitting to the NB and GB signals at a given wavevector for the **(a,b)** excitonic ($|X|^2 = 0.56$) and **(d,e)** photonic ($|X|^2 = 0.39$) detuning. **(a,d)** Show the full spectra, where the NB is fitted with the Voigt function (dashed line). **(b,e)** Zoomed spectra with examples of fitting to the weak GB signal with a Lorentzian lines (dashed lines). **(c,f)** Examples of fitting to the Bogoliubov spectrum of the GB (solid lines correspond to the reversed bare polariton dispersions).

## 6. Healing length and the crossover wavevector

The healing length of the condensate is calculated using the textbook definition $\xi = \frac{\hbar}{\sqrt{2mgn}}$, where $m$ is the polariton effective mass and $gn$ is the interaction energy extracted from the fitting of the GB excitation spectra. The calculated crossover wavevector $k_\xi = \xi^{-1}$, where the dispersion changes from linear to quadratic, is summarised in Supplementary Figure 7 for two of the detunings values probed in the experiment. The results show that our momentum-space filtering technique allows for probing the long wavevector ($k > k_\xi$) part of the elementary excitations spectrum.

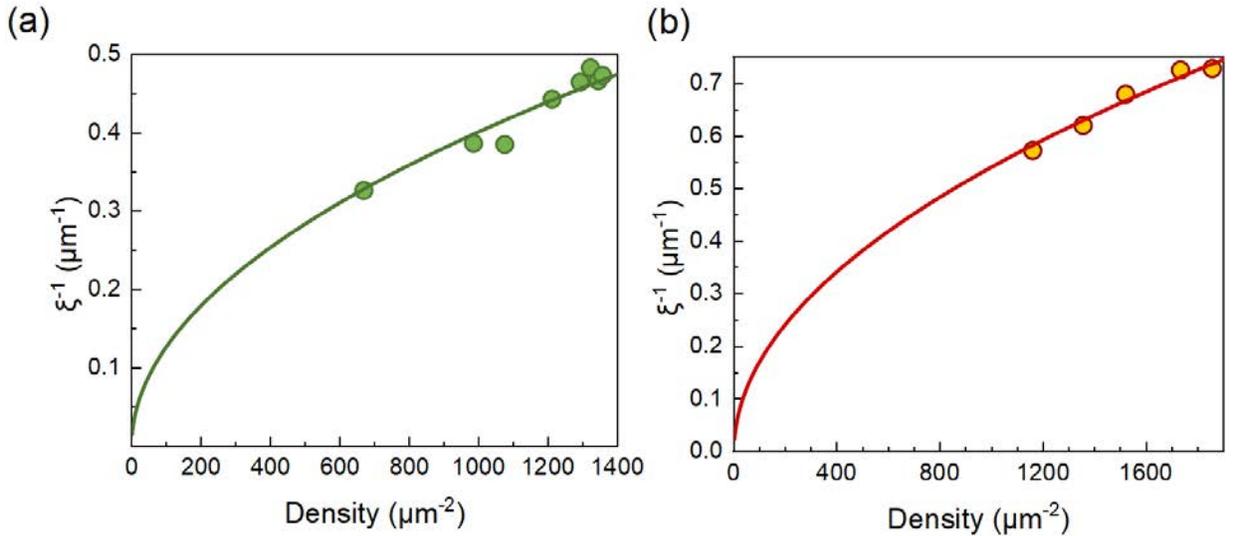

**Supplementary Figure 7** Inverse healing length of the condensate extracted for **(a)** photonic detuning ($|X|^2 = 0.39$) and **(b)** excitonic detuning ($|X|^2 = 0.56$). Solid lines are fits with a square root function.

## 7. Density dependence of the contact

As described in Methods section of the main text, the direct measurement of the GB occupation in momentum space allows one to extract the value of the Tan's contact from the $N_{GB}(k) \propto Ck^{-4}$ dependence at large k-vectors. Assuming the local density approximation (LDA) in the middle of the trap, where the condensate density is a smooth function, the proportionality coefficient should depend quadratically on the peak density $C \propto n^2$. Verification of this relation is presented in Supplementary Figure 8 at excitonic fraction of $|X|^2 \approx 0.56$. The peak density has been extracted from real-space spectra in the middle of the trap, to avoid averaging of the density with the periphery of the condensate. We note that the contact $\mathcal{C}_\infty$ measured for an atomic BEC, e.g., in Ref. [7], is defined as the limit of density distribution in momentum space rather than the occupation number. In a two-dimensional quantum gas, it is therefore related to the quantity defined above as $\mathcal{C}_\infty = CS/(2\pi)^2$, where S is the area of the condensate in real space. Both $\mathcal{C}_\infty$ and $C$ exhibit quadratic dependence on the peak density in the LDA.

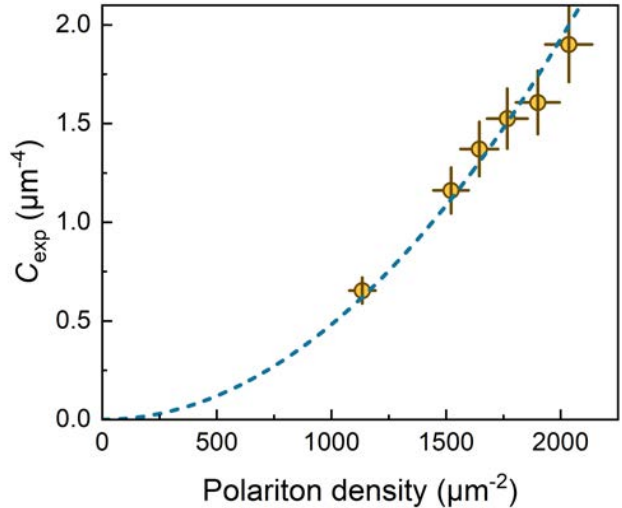

**Supplementary Figure 8** Experimentally determined values of the contact $C_{exp}$ as a function of the peak polariton density determined from the real space PL spectrum. Dashed line indicates a quadratic fit to the experimental data.

## 8. Polaritons vs massive bosons: influence of non-parabolicity of the polariton dispersion

When probing the elementary excitations of the condensate in a large momentum limit, it is important to understand to what extent the non-parabolicity of the lower polariton dispersion, $E_{LP}(k)$, which may become prominent in this limit, modifies the expressions for the energy of the Bogoliubov excitations used in the main text. We recall that the energy of the lower polariton branch can be derived from the standard coupled exciton-photon model[10], and takes the form:

$$E_{LP}(k) = \frac{1}{2}\left[E_c(k) + E_X - \sqrt{\Delta^2(k) + E_R^2}\right], \quad (S1)$$

where $E_c(k) = E_c(0) + \hbar^2 k^2/2m_c$ is the cavity photon dispersion, $m_c$ is the effective mass of the cavity photon, $E_X$ is the exciton energy (assumed constant), $\Delta(k) = E_c(k) - E_X$ is the exciton-photon detuning ranging from positive (excitonic) to negative (photonic), and $E_R = 2\hbar\Omega$ is the Rabi splitting at zero detuning.

Taking into account the polariton energy in momentum space, $E_{LP}(k)$, the energy of the elementary excitations is expressed as follows:

$$\epsilon^{LP}(k) = \sqrt{T_{LP}(k)}\sqrt{T_{LP}(k) + 2\mu}, \quad (S2)$$

where $T_{LP}(k) = E_{LP}(k) - E_{LP}(0)$, and $\mu = gn$. We note that the standard expression for the Bogoliubov dispersion in equilibrium (Eq. (2) in the main text, Methods) is recovered from Eq. (S2) by replacing the non-parabolic kinetic energy of the lower polariton $T_{LP}(k)$ with the parabolic (effective mass) approximation $T_{LP}(k) \to E(k) = \hbar^2 k^2/2m$, where $m$ is the effective mass of the lower polariton, as defined in the main text. The crossover wavevector, which is defined by the healing length as the value of $k_\xi = \xi^{-1}$, at which the transition from the phonon to the free-particle behaviour of elementary excitations occurs, i.e. $T_{LP}(k_\xi) = \mu$. Just as in the case of massive bosons, the large wavevector limit is then defined as $k \gg k_\xi$, or equivalently $T_{LP}(k) \gg \mu$.

The expression for the amplitude of the elementary excitations can then be written as

$$u_k^{LP}, v_{-k}^{LP} = \pm\frac{1}{\sqrt{2\epsilon^{LP}(k)}}\sqrt{T_{LP}(k) + \mu \pm \epsilon^{LP}(k)}, \quad (S3)$$

and the occupation of the ghost branch is:

$$N_{GB}^{LP}(k) = |v_{-k}^{LP}|^2 = \frac{1}{2}\frac{1 + \mu/T_{LP}(k)}{\sqrt{1 + 2\mu/T_{LP}(k)}} - \frac{1}{2}, \quad (S4)$$

where the ratio $\mu/T_{LP}(k)$ defines the ratio of the wavevectors $k_\xi^2/k^2 = \mu/T_{LP}(k)$ for the case of non-parabolic LP dispersion. In the limit of large momenta, $T_{LP}(k) \gg \mu$, one can expand Eq. (S4) in the powers of the small parameter $\mu/T_{LP}(k) \ll 1$. Omitting the higher-order terms allows us to obtain the asymptotic behaviour at large momenta:

$$N_{GB}^{LP}(k) \xrightarrow{k\xi \gg 1} \frac{\mu^2}{4T_{LP}^2(k)} = \frac{g^2 n^2}{4[E_{LP}(k) - E_{LP}(0)]^2}. \quad (S5)$$

This function tends to a non-zero, constant value at very large momenta:

$$N_{GB}^{LP}(k) \xrightarrow{k \to \infty} \frac{g^2 n^2}{4[E_X - E_{LP}(0)]^2} = \frac{g^2 n^2}{\left[\sqrt{\Delta^2(0) + E_R^2} - \Delta(0)\right]^2} \quad (S6)$$

which differs from the corresponding behaviour for a massive boson:

$$N_{GB}(k) \xrightarrow{k\xi \gg 1} \frac{\mu^2}{4T^2(k)} = \frac{g^2 n^2}{4|C|^4[E_c(k) - E_c(0)]^2} = \frac{m^2 g^2 n^2}{\hbar^4 k^4} \propto k^{-4}, \quad (S7)$$

where $|C|^2 = 1 - |X|^2$ is the photonic Hopfield coefficient (photon fraction) at $k = 0$. However, for the range of the wavevectors and detunings probed in the experiment, the discrepancy between the asymptotic behaviour of $N_{GB}(k)$ and $N_{GB}^{LP}(k)$ given by Eqs. (S5) and (S7), respectively, is small. The long-scale and the experimentally relevant shorter scale ($k^2/k_\xi^2 > 1$) behaviour of the two functions is illustrated in the Supplementary Figures 9-11, with $k_\xi$ defined by the condition $T_{LP}(k_\xi) = \mu$.

We note that the departure of $N_{GB}^{LP}(k)$ from $\propto k^{-4}$ asymptote at large momenta due to the non-parabolicity of the polariton dispersion cannot explain the observed behaviour for photonic detunings, since $N_{GB}^{LP}(k) \to N_{GB}(k)$ for $k^2/k_\xi^2 > 1$ as $|C|^2 \to 1$. As shown in Supplementary Figure 10, the fit to the $\propto k^{-4}$ asymptote should improve for photonic detunings, which is contrary to what is observed in the experiment (cf. Figure 5e in the main text, the highest density data shown in light green). As noted in Methods of the main text, this effect also does not affect the fitting of the experimentally measured ghost branch presented in Section 5 above, since the actual, experimentally measured lower polariton dispersion $E_{LP}(k)$ was used in the fitting, rather than its parabolic (effective mass) approximation.

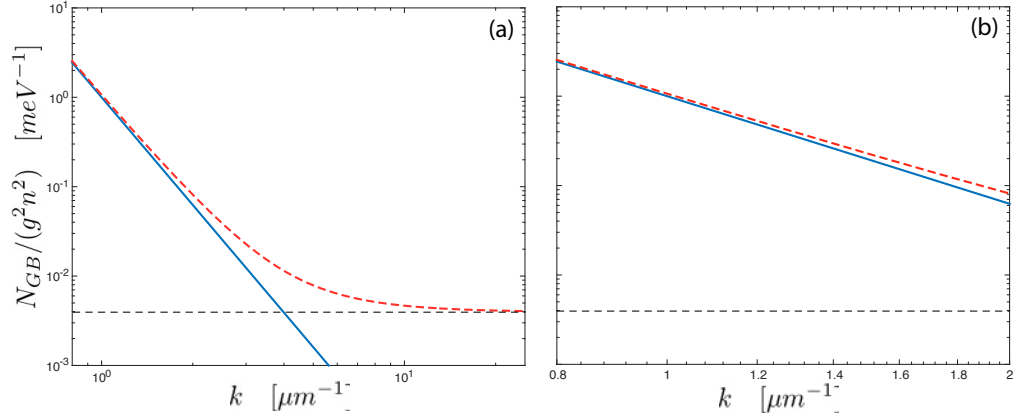

**Supplementary Figure 9** Comparison between the asymptotic behavior of $N_{GB}^{LP}(k)$ given by Eq. (S5) (red dashed), and the asymptotic behavior of $N_{GB}(k) \propto k^{-4}$ given by Eq. (S7) (blue) at (a) very long, $k^2/k_\xi^2 \gg 1$, and (b) shorter, $k^2/k_\xi^2 > 1$, ranges of the wavevector for $\Delta = 0\ meV$, $|X|^2 = 0.5$. The horizontal black dashed line is given by Eq. (S6).

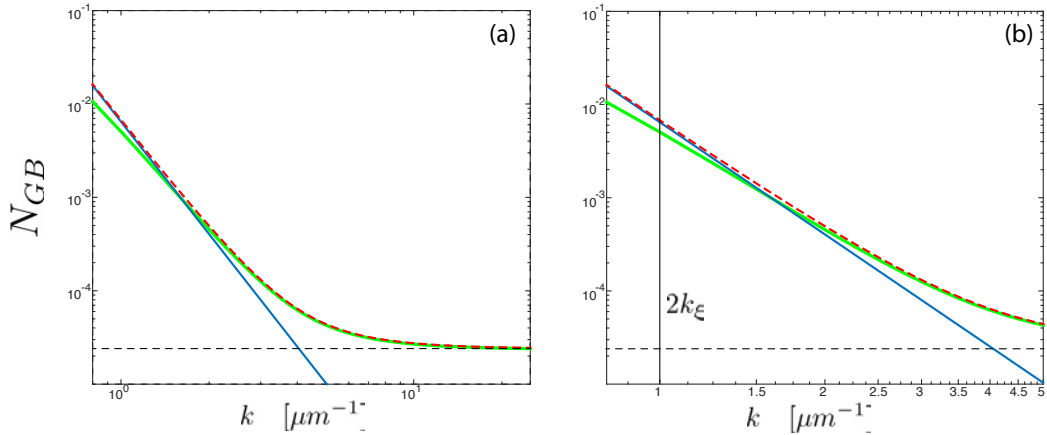

**Supplementary Figure 10** Comparison of the behaviour of $N_{GB}^{LP}(k)$ given by Eq. (S4) (light green), its asymptotic behaviour given by Eq. (S5) (red dashed), and the asymptotic behaviour of $N_{GB}(k) \propto k^{-4}$ given by Eq. (S7) (blue) at (a) very long, $k^2/k_\xi^2 \gg 1$, and (b) shorter, $k^2/k_\xi^2 > 1$, ranges of the wavevector for $\Delta = -3.7\ meV$, $|X|^2 = 0.39$, $k_\xi = 0.51\ \mu m^{-1}$. The horizontal black dashed line is given by Eq. (S6).

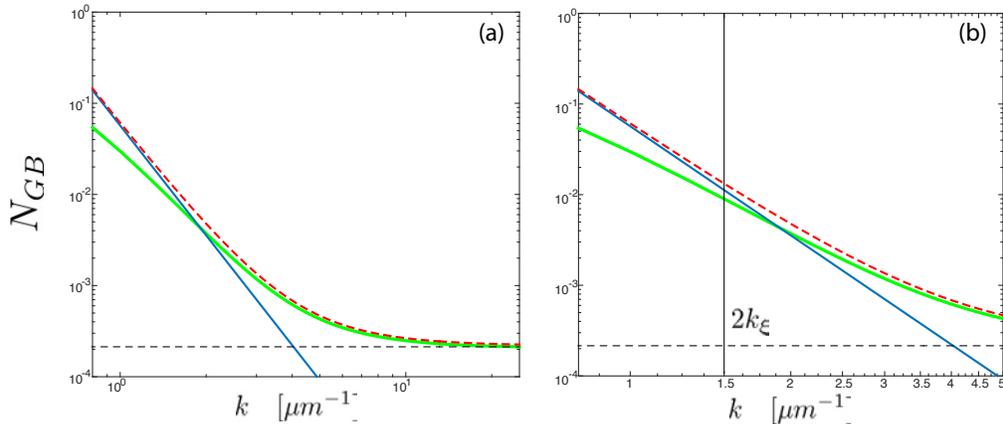

**Supplementary Figure 11** Same as Supplementary Figure 10, but for $\Delta = +1.8$ meV, $|X|^2 = 0.56$, $k_\xi = 0.76\ \mu m^{-1}$.

## 9. GB occupation at different positions on the sample (different detunings)

In Supplementary Figure 12, we present examples of data for different positions on the sample and different detunings showing that the observed behaviour is generic for the whole sample under similar experimental conditions. Supplementary Figure 12a illustrates data at a photonic detuning ($|X|^2 = 0.35$) where one can observe a gradual deviation from the power-law $k_\parallel^{-4}$ at larger densities, similar to what is presented in the main text.

In our experiment, we observe the power-law decay $k_\parallel^{-4}$ at all excitonic detunings, whithout any visible discrepancies within the probed wavevector range. An example at one of the detunings ($|X|^2 = 0.53$) is presented in Supplementary Figure 12b.

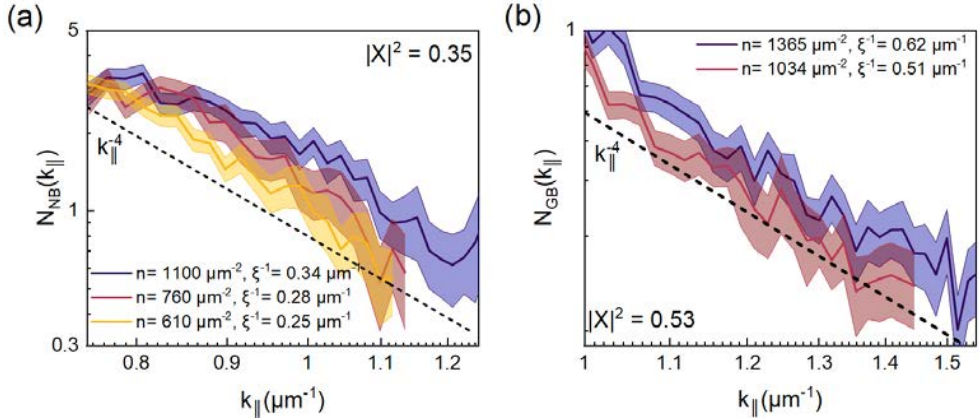

**Supplementary Figure 12** Examples of the GB occupation in momentum space at various detunings. (a) Example of a photonic detuning, where $|X|^2 = 0.35$ and densities up to $10^3\ \mu m^{-2}$ were probed. (b) Example of an excitonic detuning $|X|^2 = 0.53$, where the GB occupation follows the power law $k_\parallel^{-4}$ at all probed densities.

## 10. Raw data for the GB

In Supplementary Figure 13, we show raw data for the GB occupation presented in the main text (measured in photon counts per second), without recalculation to occupation per state in the momentum space (see Methods). One can observe that the power-law dependence is preserved.

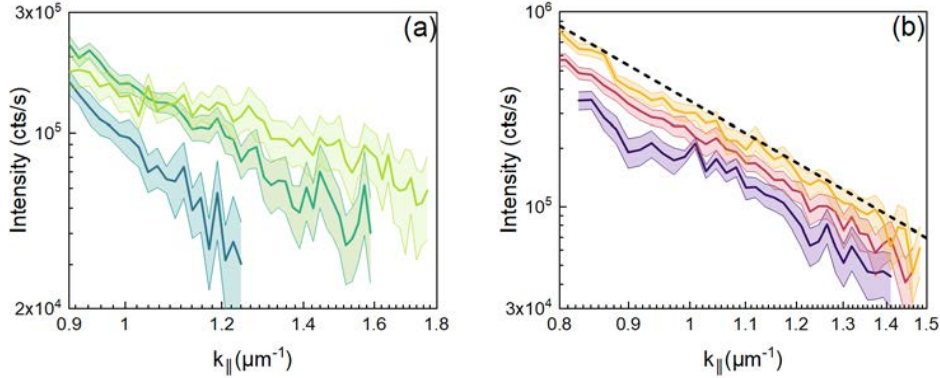

**Supplementary Figure 13** Raw data of integrated intensities for GB data presented in Fig. 5 of the main text. (a) Photonic detuning ($\Delta = -3.7\ meV, |X|^2 = 0.39$) and excitonic detuning ($\Delta = +1.8\ meV, |X|^2 = 0.56$). Dashed line in (b) is a guide to the eye for the power law $k_\parallel^{-4}$. One can observe that the power-law decay is also present in the raw data.

## 11. Domain of dynamical instability

In certain regions of system parameters, *spatially homogeneous* condensates created in the CW regime were predicted to exhibit dynamical (modulational) instability, which is driven by fluctuations of the reservoir[11,12]. This instability was experimentally observed in large, quasi-1D condensates[13]. Analytical estimate obtained using the open-dissipative Gross-Pitaevskii model[11] yields the instability domain for the *homogeneous* pump rate (i.e. the rate of reservoir replenishing): $P/P_{th} < P_{MI}/P_{th} = \gamma g_R/(g\gamma_R)$, where the threshold pump rate is defined as $P_{th} = \gamma\gamma_R/R$ (other parameters are defined in Sec. 3). The boundary of the instability region can be expressed as a function of the excitonic Hopfield coefficient:

$$P_{MI}/P_{th} = 1 + \frac{\gamma_{ph}}{\gamma_X}\frac{(1-|X|^2)}{|X|^2}, \qquad (S8)$$

where we have assumed that the decay rate of highly excitonic reservoir particles is equal to that of excitons, and $\gamma_{ph}$ is the decay rate of cavity photons. For our long-lifetime system, the ratio $\gamma_{ph}/\gamma_X \approx 10$, and the corresponding instability domain is shown in Supplementary Figure 14a. In this domain, the elementary excitations within the band of wavevectors[11] $k/k_\xi < 2\sqrt{P_{MI}/P - 1}$, shown in Supplementary Figure 14b, may grow exponentially. As pointed out in Ref[14], this means that the fluctuations become large, the adiabatic approximation for the reservoir dynamics (which is the basis for Eq. (3) in Methods) breaks down, and the population of the GB departs from the Bogoliubov prediction. Numerical simulations of the open-dissipative mean field model show deviation of the $N(k)$ from the Bogoliubov prediction and characteristic flattening of the momentum dependence[14,15], however the details of this behaviour are model-specific.

The domain of modulational instability is strongly suppressed for spatially inhomogeneous condensates[16]. However, our trapped condensates created at photonic detunings ($|X|^2 < 0.5$) approach the "flat-top", spatially homogeneous profile shapes at large pump powers and large densities (see, e.g., Supplementary Figure 5a). It is therefore possible that, as we increase the pump power, we drive the system into the modulational instability domain[14], where large deviations from the Bogoliubov distribution $N(k)$ are expected. Indeed, such deviations are observed in our experiment (see Fig. 5 of the main text and Supplementary Figure 12a). However, we do not observe strong density modulations of the condensate typically associated with the dynamical instability[13], which requires further investigation of the relevance of this regime and the corresponding mean-field models to our observations.

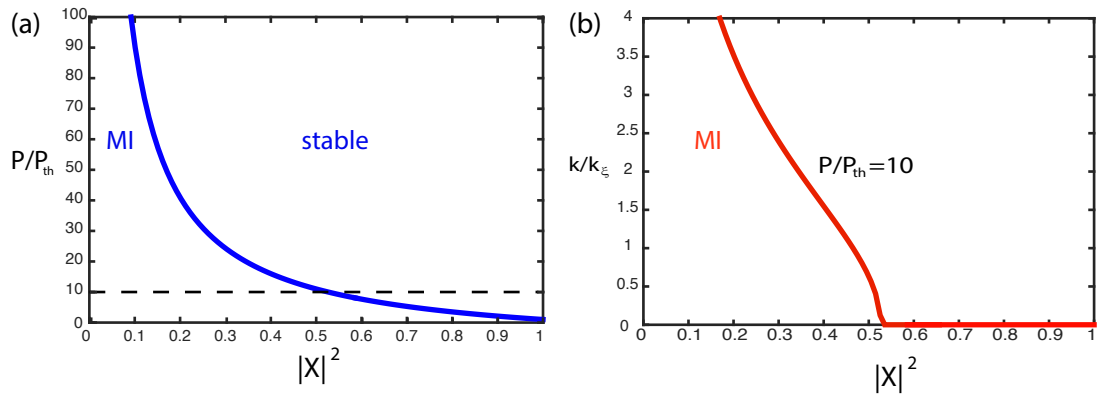

**Supplementary Figure 14** (a) Boundary of the dynamical (modulational) instability of a homogeneous CW condensate given by Eq. (S8). (b) Domain of unstable excitation wavevectors corresponding to the pump rate relative to the condensation threshold indicated in (a) by a horizontal dashed line.